\documentclass[%
 reprint,
 amsmath,amssymb,
 aps,
 prl,
]{revtex4-2}

\usepackage{graphicx}
\usepackage{dcolumn}
\usepackage{bm}
\usepackage{color}

\newcommand{\bfu}{\bm{u}}

\newcommand{\bfx}{\bm{x}}

\newcommand{\bfell}{\bm{\ell}}
\newcommand{\bfb}{\bm{b}}
\newcommand{\bfB}{\bm{B}}
\newcommand{\bfj}{\bm{J}}
\newcommand{\bfo}{\bm{\omega}}
\newcommand{\bnab}{\bm{\nabla}}

\renewcommand{\P}{\mathcal{P}}
\newcommand{\D}{\mathcal{D}}
\newcommand{\M}{\mathcal{M}}

\newcommand{\Rm}{\mathrm{Rm}}
\renewcommand{\Re}{\mathrm{Re}}

\newcommand{\V}{\mathcal{V}}
\newcommand{\Exp}[2]{\left\langle{#1}\right\rangle_{#2}}

\graphicspath{{Figures/}}

\begin{document}


\title{Local relaxation and scale-dependent alignment \\ in compressible, magnetized turbulence}

\author{James R. Beattie$^{\dagger,}$}
 \email{beattie@ias.edu \\ $^\dagger$NASA Hubble Fellow}
 \affiliation{
Department of Astrophysical Sciences, Princeton University, Princeton, 08540, NJ, USA \\
Canadian Institute for Theoretical Astrophysics, University of Toronto, Toronto, M5S3H8, ON, Canada \\
School of Natural Sciences, Institute for Advanced Study, 1 Einstein Drive, Princeton, NJ 08540, USA
}
\author{Amitava Bhattacharjee}%
\affiliation{%
Department of Astrophysical Sciences, Princeton University, Princeton, 08540, NJ, USA
}%

\date{\today}

\begin{abstract}
    Driven net- and no-net-flux MHD turbulence simulations up to $10,\!368^3$ reveal sign-mixed velocity--magnetic, velocity--vorticity, and magnetic--current aligned patches below the energy equipartition scale. The first two angles scale as $\lambda^{1/8}$ and $\lambda^{1/16}$, while the third angle varies weakly with scale. We develop and test a constant-flux transport model for departures from relaxed states, which predicts both exponents. These findings affect eddy anisotropy, reconnection-mediated turbulence onset, large-scale dynamos, and the nature of magnetized turbulence. 
\end{abstract}

\maketitle
\vspace{-5mm}
\section{Introduction}
\vspace{-2mm}
    Magnetized turbulence pervades laboratory and planetary plasmas \citep{Jones2011_planetary_dynamo_review,Brun2017_solar_dynamo_review,Tzeferacos2018_dynamo_in_the_lab}, the solar wind and magnetosheath \citep{Roberto2013_solar_turb_review,Pecora2023_relaxation_in_magnetosheath}, star-forming clouds and the interstellar medium \citep{Elmegreen2004,Klessen2011,Gaesnsler_2011_trans_ISM,Beattie2025_nature_astro}, and intergalactic and intracluster media \citep{Schmidt2021_CGM_turbulence,Mohapatra2019_turbulent_heat_flux_ICM}. Its nonlinear MHD interactions regulate cascade rates, mixing, heating, reconnection, and the turbulent transport that sustains large-scale dynamos. Alignment weakens nonlinear interactions between velocity, $\bfu$, and Alfv\'enic magnetic fluctuations, $\bfb$. Counter-propagating Elsasser fluctuations, $\bm z^\mp=\bfu\mp\bfb$, interact on $t_{\rm nl}\sim(k_{\perp}z_{\perp}^{\mp}\sin\theta)^{-1}$, so $\bfu=\pm\bfb$ suppresses the Alfv\'enic nonlinearity \citep{Dobrowolny1980_solar_wind_properties,Boldyrev2006,Mason2006_dynamical_alignment,Perez2009_dynamical_alignment_of_imbalanced_islands,Mallet2015_refined_cb,Chandran2015_alignment,Chernoglazov2021_alignment_SR_MHD,Sioulas2024_scale_dependent_alignment_solar_wind}. In guide-field turbulence, $\theta\sim\lambda/\xi$ also controls the perpendicular eddy thickness, $\lambda$, relative to its length, $\xi$ \citep{Mallet2016_measures_anisotropy_intermittency,Mallet2017_anisotropy}. A scale-independent energy flux $\epsilon\sim\delta u_\lambda^3\theta/\lambda$ then gives the dynamic-alignment prediction $\theta\sim\lambda^{1/4}$ and a $k_\perp^{-3/2}$ spectrum \citep{Boldyrev2006,Mason2006_dynamical_alignment,Perez2012_energy_spectrum_strong_MHD,Chandran2015_alignment}, with implications for reconnection-mediated turbulence \citep{Loureiro2017_reconnection_in_turbulence,Comisso2018_MHD_turbulence_plasmoid_regime,Dong2018_role_of_plasmoid_instability,Dong2022_reconnection_mediated_cascade}.

    However, alignment also has a broader interpretation, in that it is a signature of plasma relaxation \citep{Taylor1974_relaxation,Matthaeus2008_rapid_alignment,Servidio2008_depression_nonlinearity,Banerjee2023_relaxed_states,Pecora2023_relaxation_in_magnetosheath}. Relaxed MHD states are configurations in which nonlinear forces cancel, or become gradients, unable to drive further turbulent or nonlinear transfer. Three alignments are relevant when considering the MHD model. Alfv\'enic alignment (Alfv\'enization), $\bfu\parallel\bfB$, weakens induction and shear-Alfv\'en interactions; Taylor alignment (Taylorization), $\bfj\parallel\bfB$, weakens the Lorentz force; and Beltrami alignment (Beltramization), $\bfu\parallel\bfo$, makes the inertial term gradient-like. Here $\bfj=\bnab\times\bfB$ is the current density and $\bfo=\bnab\times\bfu$ is the vorticity. Each alignment corresponds directly to an ideal helicity invariant: cross helicity, $\bfu\cdot\bfB$; current helicity, $\bfB\cdot\bfj$; or kinetic helicity, $\bfu\cdot\bfo$. Classically, these aligned states arise by finding the global minimum-energy state of the flow, constrained by the helicities \citep{Taylor1974_relaxation,Matthaeus2008_rapid_alignment,Servidio2008_depression_nonlinearity,Banerjee2023_relaxed_states,Pecora2023_relaxation_in_magnetosheath}. However, both in simulation and in the Earth's magnetosheath the alignments are localized and patchy across space \citep{Servidio2008_depression_nonlinearity,Hosking2021_reconnection_controlled_decay,Pecora2023_relaxation_in_magnetosheath}. 

    \begin{figure*}
        \centering
        \includegraphics[width=0.93\linewidth]{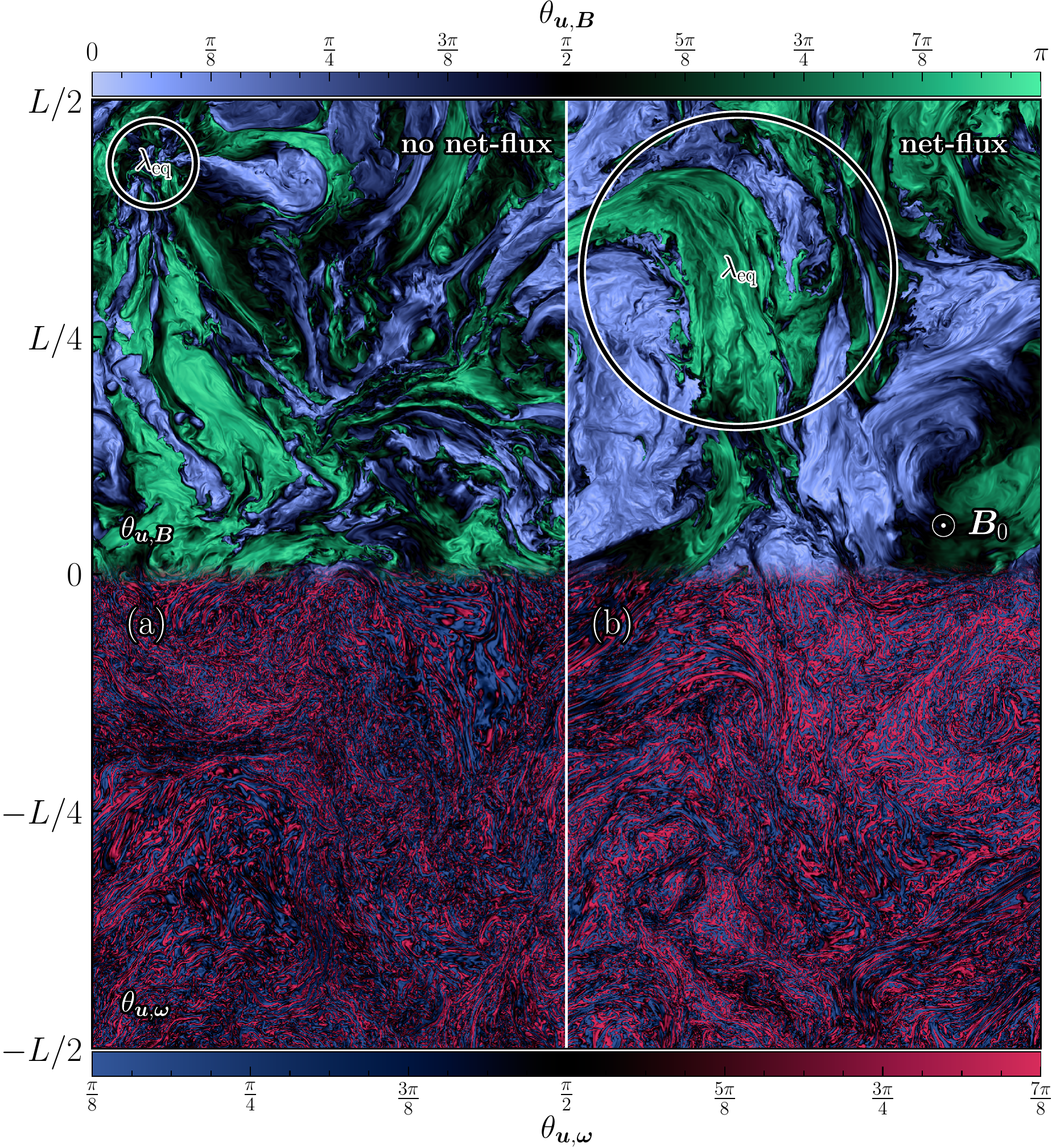}
        \vspace{-2mm}
        \caption{\textbf{Local alignment in no-net-flux and net-flux turbulence.} Composite full-domain slices for \textbf{(a)} the no-net-flux simulation and \textbf{(b)} the net-flux guide-field simulation. In each column, the top half shows $\theta_{\bfu,\bfB}$ and the bottom half shows $\theta_{\bfu,\bfo}$ from the same slice; in \textbf{(b)}, $\bfB_0$ is directed out of the plane. The color encodes the local angle from parallel through perpendicular to anti-parallel alignment, with the lower $\theta_{\bfu,\bfo}$ color scale clipped at $\pi/8$ and $7\pi/8$ for contrast. Dark interfaces mark nearly orthogonal regions where the corresponding cross products, $\bfu\times\bfB$ or $\bfu\times\bfo$, are largest. White circles have radius $\lambda_{\rm eq}$, the energy-equipartition scale where the local Alfv\'en and eddy times are equal, with $\lambda_{\rm eq}/L=0.09\pm0.01$ in \textbf{(a)} and $0.33\pm0.06$ in \textbf{(b)}.}
        \vspace{-5mm}
        \label{fig:v_b_theta}
    \end{figure*}

    Recent work by \citet{Banerjee2023_relaxed_states} describes a theory for the relaxation of scale-dependent helicity and invariant correlations, named the principle of vanishing nonlinear transfer (PVNLT). In PVNLT, a relaxed state has vanishing nonlinear transfer of the invariant correlations (see Supplemental Material~\citep{SupplementalMaterial}, Sec.~III, for more details). The limit in which the two-point helicity correlation functions become stationary gives
    \begin{align}
        \bfu\times\bfB
        &\simeq0
        \quad\Rightarrow\quad
        \bfu=\kappa\bfB,
        \qquad
        \bfo=\kappa\bfj,
        \label{eq:intro_pvnlt_alfvenic_branch}\\
        (1-\kappa^2)\bfj\times\bfB
        &=
        \bnab(P_T+\phi_0),
        \label{eq:intro_pvnlt_taylor_balance}
    \end{align}
    where $P_T$ is the total pressure and $\phi_0$ is a problem-dependent scalar potential. Thus $\kappa=\pm1$ gives the Alfv\'enic states, $\bfu=\pm\bfB$, while the magnetically dominated or pressure-balanced limits give Taylor alignment, $\bfj\parallel\bfB$; when the Lorentz term is force-free or gradient-balanced, the same relaxation condition depletes the non-gradient part of $\bfu\times\bfo$, giving Beltrami alignment. In this study we show that departures from these relaxed states are themselves transferred through the cascade, which we use to model the scale-dependent alignment structure functions.
        
    We show numerically and analytically that below the energy-equipartition scale $\lambda_{\rm eq}$, where Alfv\'en and eddy times are equal and rapid relaxation can occur \citep{Matthaeus2008_rapid_alignment,Servidio2008_depression_nonlinearity}, eddies remain close to local relaxed states given by Eqs.~\eqref{eq:intro_pvnlt_alfvenic_branch} and \eqref{eq:intro_pvnlt_taylor_balance}. Here $\lambda_{\rm eq}/L=0.09\pm0.01$ for no-net-flux and $0.33\pm0.06$ for net-flux. However, the large scales are not perfectly relaxed and they pass departures from Alfv\'enic and Beltrami states to smaller scales with an approximately constant flux, which we verify directly with shell-transfer measurements. For the measured subsonic scaling $u_k^2\sim k^{-3/2}dk$ \citep{Beattie2025_nature_astro}, constant downscale transport of the Alfv\'enic departure $\D_A\sim\theta_{\bfu,\bfB}^2$ predicts $\theta_{\bfu,\bfB}\sim\lambda^{1/8}$. Because kinetic helicity is signed and globally cancels, the leading sign-even Beltrami departure is $\D_B^2\sim\theta_{\bfu,\bfo}^4$, giving $\theta_{\bfu,\bfo}\sim\lambda^{1/16}$. We test these predictions with two extreme-resolution, low plasma beta, $\beta$, compressible MHD simulations performed on SuperMUC-NG, using more than $1.6\times10^8$ core-hours: (1) a $10,\!080^3$ no-net-flux fluctuation-dynamo simulation \citep{Beattie2025_nature_astro} and (2) a $10,\!368^3$ strong-guide-field simulation. Both have $\Re\sim\Rm\gtrsim10^6$ and nearly three decades in resolved, inertial scales, spanning a large-scale, kinetically dominated supersonic cascade with $u_k^2\sim k^{-2}dk$ and a subsonic, magnetically dominated cascade with $u_k^2\sim k^{-3/2}dk$ \citep{Beattie2025_nature_astro}. This plasma parameter range spans most phases of interstellar turbulence, weakly compressible space plasmas, and many astrophysical laboratory experiments \citep{Ferriere2020_reynolds_numbers_for_ism,Bandyopadhyay2025_MMS_compared_with_10k,Tzeferacos2018_dynamo_in_the_lab,Merlini2026_lab_shock_turbulence}.
\vspace{-5mm}
\section{Numerical simulations}
\vspace{-2mm}
    We solve isothermal ideal MHD in a periodic domain with a hybrid-precision \textsc{flash} implementation and a positivity-preserving MUSCL-Hancock HLL5R scheme \citep{Fryxell2000,Dubey2008,Bouchut2010,Waagan2011,Federrath2021}. A finite-correlation-time, nonhelical Ornstein--Uhlenbeck force drives both simulations at $kL=2$ \citep{Eswaran1988_forcing_numerical_scheme,Schmidt2009,Federrath2010_solendoidal_versus_compressive,Federrath2022_turbulence_driving_module}. Their stationary states have $\M\approx4$, $\beta=1/2$, and Alfv\'enic Mach number $\approx2$ (defined using the rms field for the no-net-flux run and $B_0$ for the guide-field run); the latter has $\delta B/B_0\approx1$ \citep{Beattie2020c}. The End Matter section ``Numerical details and Reynolds estimate'' gives the initial conditions and effective $\Re\sim\Rm\sim10^6$.
    
    The simulations have zero signed system-scale fluctuating helicities,
    \begin{align}
        \left\langle\bfu\cdot\bfB\right\rangle_{\V}\approx 0,\qquad
        \left\langle\bfj\cdot\bfB\right\rangle_{\V}\approx 0,\qquad
        \left\langle\bfu\cdot\bfo\right\rangle_{\V}\approx 0,
        \label{eq:global_zero_helicities}
    \end{align}
    within statistical uncertainty. This excludes a single global relaxed state but permits sign-mixed local relaxation, for example,
    \begin{align}
        \bfu\simeq \kappa(\bfx)\bfB,
        \qquad
        \left\langle\kappa(\bfx) B^2\right\rangle_{\V}=0,
        \label{eq:sign_mixed}
    \end{align}
    where $\kappa(\bfx)$ is signed. Such alignment can deplete induction and Lorentz forces and make the inertial term gradient-like while canceling in volume averages.

    \begin{figure*}
        \centering
        \includegraphics[width=0.82\linewidth]{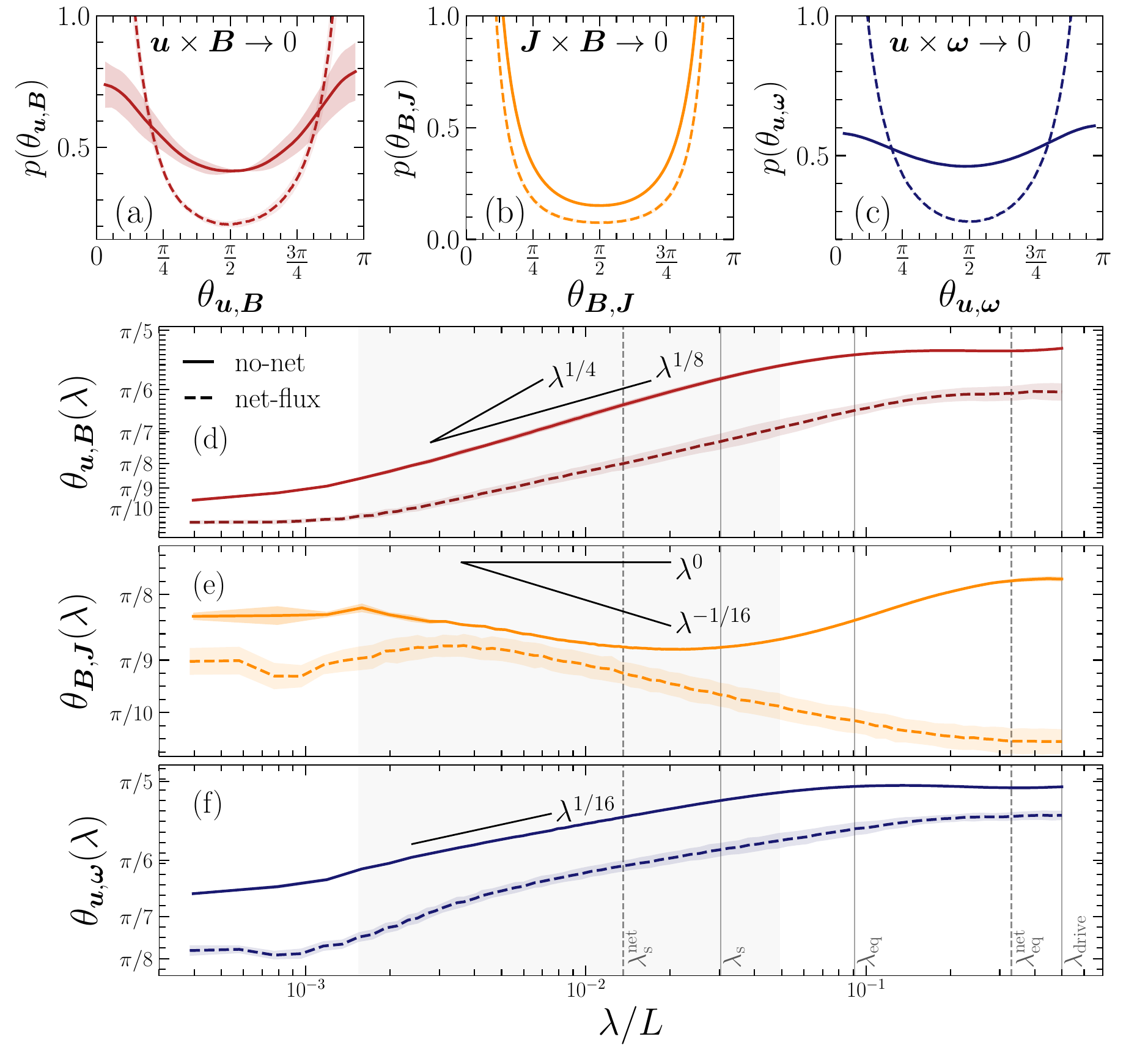}
        \vspace{-0.7cm}
        \caption{\textbf{Global and scale-dependent alignment.} \textbf{(a)--(c)} PDFs of $\theta_{\bfu,\bfB}$, $\theta_{\bfB,\bfj}$, and $\theta_{\bfu,\bfo}$, with peaks near parallel and anti-parallel Alfv\'enic, Taylor, and Beltrami states \citep{Stribling1991_relaxation_processes,Servidio2008_depression_nonlinearity,Banerjee2023_relaxed_states,Pecora2023_relaxation_in_magnetosheath}. \textbf{(d)--(f)} Alignment structure functions of projected increments versus the strong-cascade separation $\lambda$: $\lambda=|\bfell|$ in the no-net-flux run and $\lambda=|\bfell_\perp|$ in the net-flux guide-field comparison, with increments projected perpendicular to the local two-point mean magnetic field before the angle is formed. Solid curves are no-net-flux statistics; dashed curves are net-flux guide-field statistics. Shaded bands show realization-to-realization $1\sigma$ scatter; vertical grey lines mark $\lambda_s$, $\lambda_{\rm eq}$, and the driving scale as labeled. Below $\lambda_{\rm eq}$, $\theta_{\bfu,\bfB}\sim\lambda^{1/8}$ in both simulations and $\theta_{\bfu,\bfo}\sim\lambda^{1/16}$. The $\theta_{\bfB,\bfj}$ variation is small, $\Delta\theta_{\bfB,\bfj}\le0.064\,{\rm rad}$ ($3.7^\circ$) below $\lambda_{\rm eq}$, which is the sense in which the magnetic--current alignment is nearly scale independent; the guide-field curve remains compatible with a weak $\lambda^{-1/16}$ trend. The $\lambda^{1/4}$ guide line is the dynamic-alignment prediction \citep{Boldyrev2006,Mason2006_dynamical_alignment}.}
        \label{fig:alignment_variables}
        \vspace{-0.5cm}
    \end{figure*}
\vspace{-5mm}
\section{Patchy alignment and local relaxation}
\vspace{-2mm}
    Fig.~\ref{fig:v_b_theta} shows that the compressible turbulence (left, no-net-flux; right, net-flux) organizes into extended patches with $\bfu\propto\pm\bfB$ (top) and $\bfu\propto\pm\bfo$, separated by thin orthogonal interfaces, where the corresponding nonlinearities are largest. The white circles mark $\lambda_{\rm eq}$, the scale below which the cascade becomes magnetically dominated. Supplemental Material Fig.~S1 zooms into the guide-field $\theta_{\bfu,\bfB}$ slice and overlays the in-plane $\bfu_\perp$ and $\bfB_\perp$ streamlines inside an aligned patch and across an interface. The PDFs in Fig.~\ref{fig:alignment_variables}(a)--(c) show that the same local organization also includes $\bfB\propto\pm\bfj$. We interpret the measured angles as distances from local relaxed states associated with the nonlinearities in the fluid equations, not only as geometric factors in an eddy-turnover time \citep{Boldyrev2006,Beattie2025_nature_astro}. This interpretation is motivated by PVNLT \citep{Banerjee2023_relaxed_states}, where relaxed MHD states have vanishing nonlinear transfer of ideal invariant correlations. The nonlinear terms need not each vanish separately: they may vanish or be exactly balanced by gradients. Alfv\'enic, Taylor, and Beltrami alignments are the measurable limiting forms of this relaxation, with the observed $\bfB$--$\bfj$ alignment signaling depleted or pressure-balanced Lorentz forces in both simulations.
\vspace{-5mm}
\section{Scale-dependent alignment}
\vspace{-2mm}
    We measure first-order alignment structure functions at separation $\lambda$ (definitions in the End Matter section ``Definitions of scale-dependent alignment structure functions''). Figure~\ref{fig:alignment_variables}(d)--(f) shows the isotropic no-net-flux and perpendicular guide-field statistics defined in its caption. The sonic scale $\lambda_s$, where $\M(\lambda_s)=1$, is $\lambda_s/L=0.031\pm0.002$ (no-net-flux) and $0.0136\pm0.0009$ (net-flux) \citep{Federrath2021,Beattie2025_nature_astro}. At the energy-equipartition scale $\lambda_{\rm eq}$, the scale-dependent Alfv\'en and eddy times are equal; smaller scales are magnetically dominated \citep{Beattie2025_nature_astro}.

    The scale-dependent range in the $\theta_{\bfu,\bfB}$ and $\theta_{\bfu,\bfo}$ structure functions begins at $\sim\lambda_{\rm eq}$, which also corresponds to the characteristic size of the aligned patches in Fig.~\ref{fig:v_b_theta}. Below this scale, the no-net-flux run (solid lines) gives
    \begin{align}
        \theta_{\bfu,\bfB}(\lambda)\sim\lambda^{1/8}, \qquad
        \theta_{\bfu,\bfo}(\lambda)\sim\lambda^{1/16},
    \end{align}
    while $\theta_{\bfB,\bfj}(\lambda)$ remains strongly aligned with only weak scale variation. The net-flux guide-field statistics (dashed lines) show the same shallow $\theta_{\bfu,\bfB}\sim\lambda^{1/8}$ trend, offset to smaller angles, a comparable $\theta_{\bfu,\bfo}\sim\lambda^{1/16}$ trend, and the same $\theta_{\bfB,\bfj}$ variation compatible with $\lambda^{-1/16}$. In both simulations, the Alfv\'enic exponent is shallower than the strong-guide-field dynamic-alignment prediction $\theta_{\bfu,\bfB}\sim\lambda^{1/4}$ \citep{Boldyrev2006,Mason2006_dynamical_alignment,Perez2009_dynamical_alignment_of_imbalanced_islands}. Since the same $\lambda^{1/8}$ scaling appears in both simulations when measured on the appropriate cascade scale, it is unlikely to be a peculiarity of the no-net-flux fluctuation dynamo. The $\theta_{\bfu,\bfo}\sim\lambda^{1/16}$ scaling is, to our knowledge, the first measurement of scale-dependent Beltramization in MHD turbulence. 

    Because these structure functions are insensitive to the sign of the local alignment, they should be interpreted as average distances from the nearest parallel or anti-parallel PVNLT relaxed state. We established the sign-mixed character of those states by the real-space patches (Fig.~\ref{fig:v_b_theta}), the angle PDFs (Fig.~\ref{fig:alignment_variables}(a)--(c)), and the vanishing signed helicity averages in Eq.~\eqref{eq:global_zero_helicities}, while the structure functions show how the corresponding unsigned departures decrease with scale.

    \begin{figure}
        \centering
        \includegraphics[width=\linewidth]{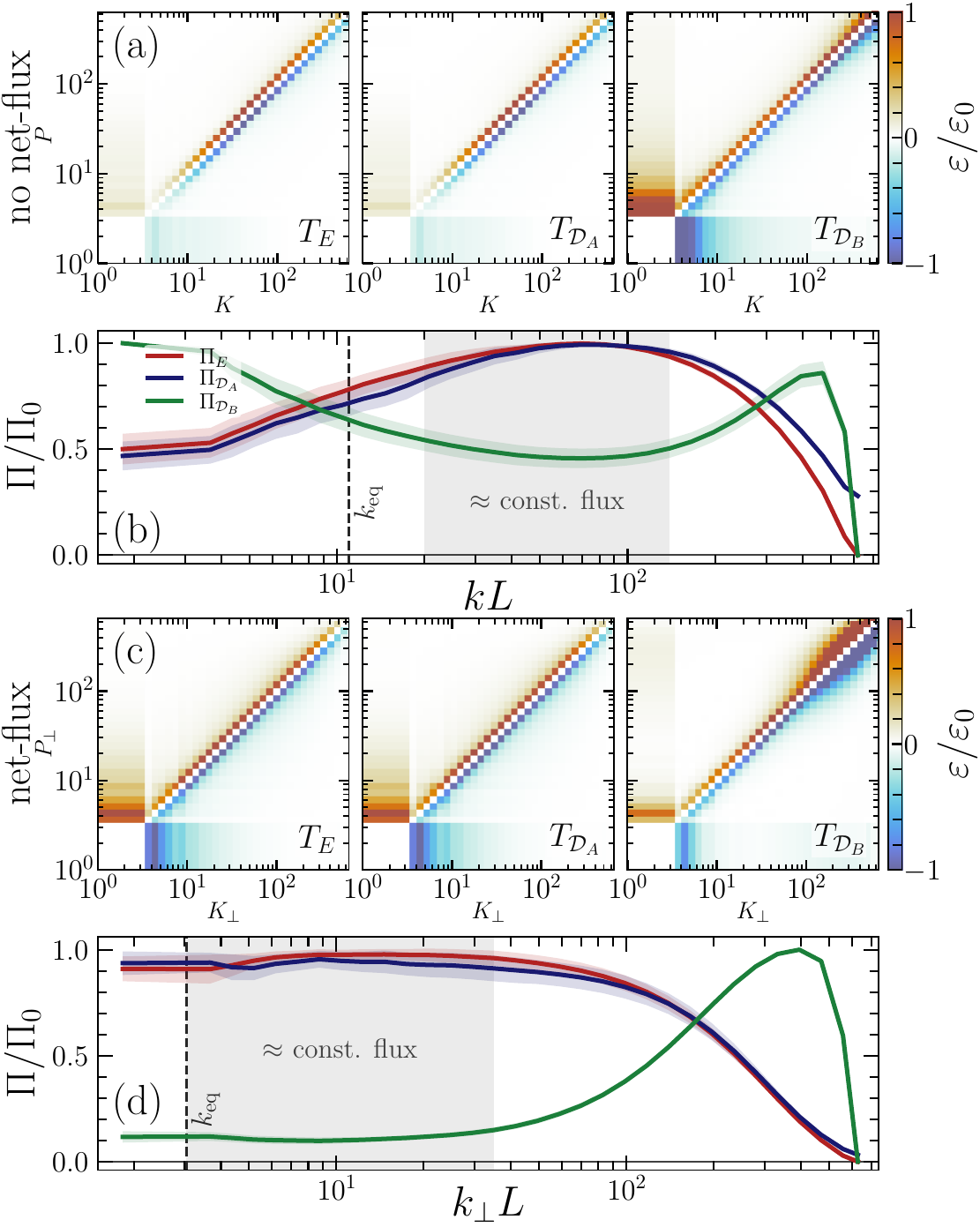}
        \vspace{-8mm}
        \caption{\textbf{Shell transfers and departure fluxes.} \textbf{(a),(c)} Shell-to-shell matrices $T(K|P)$, from source shell $P$ to target shell $K$, for total energy, $T_E$, Alfv\'enic departure, $T_{\mathcal{D}_A}$, and Beltrami departure, $T_{\mathcal{D}_B}$, in the no-net-flux and net-flux simulations. Matrix colors are normalized by $\varepsilon_0$, the characteristic shell-transfer amplitude in the corresponding constant-flux interval. \textbf{(b),(d)} Downscale fluxes normalized by $\Pi_0=\max|\Pi|$ for each realization: $\Pi_E$ (red), $\Pi_{\mathcal{D}_A}$ (blue), and $\Pi_{\mathcal{D}_B}$ (green). Shaded bands show $1\sigma$ realization scatter; grey bands mark the approximate constant-flux intervals; dashed lines mark $k_{\rm eq}$. No-net-flux transfers use isotropic shells, $K$, $P$, and $kL$; guide-field transfers use perpendicular shells, $K_\perp$, $P_\perp$, and $k_\perp L$.}
        \label{fig:alfvenic_flux}
        \vspace{-6mm}
    \end{figure}

    These three measurements are captured by small departures from relaxed states. Consider the velocity and magnetic increment amplitudes scale as $\delta u_\lambda\sim\delta B_\lambda$, e.g., on scales below $\lambda_{\rm eq}$. Using increment vectors $\bm{\delta}\bfu_\lambda$ and $\bm{\delta}\bfB_\lambda$, the Alfv\'enic departure $\D_A$ is the positive quadratic distance from $\bfu=\pm\bfB$,
    \begin{align}
        \D_A(\lambda)=
        \min_{\sigma=\pm1}
        \frac{|\bm{\delta}\bfu_\lambda-\sigma\bm{\delta}\bfB_\lambda|^2}
        {|\bm{\delta}\bfu_\lambda|^2+|\bm{\delta}\bfB_\lambda|^2}
        \simeq \frac{1}{2}\theta_{\bfu,\bfB}^2(\lambda),
        \label{eq:DA_main}
    \end{align}
    where the last relation assumes nearly equipartitioned, small-angle increments. This difference norm is natural because Alfv\'enic relaxation requires equality, $\bm{\delta}\bfu_\lambda=\pm\bm{\delta}\bfB_\lambda$, not only parallelism. A scale-independent flux $\varepsilon_{\D_A}$ for carrying this first moment downscale, $\varepsilon_{\D_A}\sim \delta u_\lambda^3\D_A/\lambda$, gives
    \begin{align}
        \theta_{\bfu,\bfB}(\lambda)
        \propto \lambda^{(1-3\zeta)/2}
        \simeq \lambda^{1/8}.
    \end{align}
    We test this prediction using the shell-to-shell departure transfer $T_{\mathcal{D}_A}(K|P)$ and its cross-scale flux $\Pi_{\mathcal{D}_A}$ (definitions in the End Matter section ``Departure transfer diagnostics''). Figure~\ref{fig:alfvenic_flux} shows that $\Pi_{\mathcal{D}_A}$ closely tracks the energy flux, becomes approximately constant near $k_{\rm eq}$, and is dominated by interactions between neighboring shells. We also compute total-energy transfers as references. The fields are filtered and downsampled before analysis; normalization, shell definitions, and conservative transfer operators are specified in End Matter. The shared constant-flux interval directly ties departure transport to the energy cascade.

    For $\bfu$--$\bfo$, the analogous signed density is kinetic helicity. The End Matter section ``Cross and kinetic helicity transport in the barotropic and force-free limits,'' Eq.~\eqref{eq:app_kinhel_balance}, shows that the kinetic helicity is made non-conservative by a non-gradient Lorentz force, but the observed Taylorization strongly suppresses that source. The remaining issue is sign: in the present nonhelical system the signed first moment can cancel between Beltrami and anti-Beltrami patches. The lowest useful sign-even Beltrami constraint is therefore the second moment of the Beltrami departure. With vorticity increment $\bm{\delta}\bfo_\lambda$,
    \begin{align}
        \D_B(\lambda)=
        1-\frac{|\bm{\delta}\bfu_\lambda\cdot\bm{\delta}\bfo_\lambda|}
        {|\bm{\delta}\bfu_\lambda||\bm{\delta}\bfo_\lambda|}
        \simeq\frac{1}{2}\theta_{\bfu,\bfo}^2(\lambda),
        \label{eq:DB_main}
    \end{align}
    which is angular because Beltrami states require proportionality between $\bfu$ and $\bfo$, whose amplitudes and dimensions differ. The amplitude-weighted transfer diagnostic used to measure this transport is defined in the End Matter section ``Departure transfer diagnostics,'' Eqs.~\eqref{eq:app_beltrami_vector_departure} and \eqref{eq:app_beltrami_flux}. The approximately constant $\Pi_{\mathcal{D}_B}(k)$ intervals in Fig.~\ref{fig:alfvenic_flux} support a scale-independent rate $\varepsilon_{\D_B}$ for carrying this second moment downscale, $\varepsilon_{\D_B}\sim\delta u_\lambda^3\D_B^2/\lambda$, which gives
    \begin{align}
        \theta_{\bfu,\bfo}(\lambda)
        \propto \lambda^{(1-3\zeta)/4}
        \simeq \lambda^{1/16}.
    \end{align}
    Thus the different powers of the angle are not introduced symmetrically. The Alfv\'enic case has a positive first-moment departure inherited from the energy--cross-helicity square, while the kinetic-helicity case is signed and globally canceled, making a departure correlation the leading sign-even object. Supplemental Material, Sec.~IV, gives a more detailed exponent derivation.
\vspace{-5mm}
\section{Discussion \& Conclusions}
\vspace{-2mm}
    Driven compressible MHD turbulence self-organizes into sign-mixed relaxed patches below $\lambda_{\rm eq}$, analogous to cross-helical patches in the solar wind and magnetosheath \citep{Matthaeus2008_rapid_alignment,Osman2011_solar_wind_alignment,Pecora2023_relaxation_in_magnetosheath}. The measured shell transfers support a constant, local cascade of departures from these states: with $\delta u_\lambda\sim\lambda^{1/4}$, $\D_A\sim\theta_{\bfu,\bfB}^2$ and $\D_B^2\sim\theta_{\bfu,\bfo}^4$ yield the observed $\lambda^{1/8}$ Alfv\'enic and $\lambda^{1/16}$ Beltrami exponents. Strong Taylorization suppresses the Lorentz-force source in the kinetic-helicity evolution, enabling conservative Beltrami-departure transport.

    Scale-dependent alignment sets eddy anisotropy: $\lambda/\xi\sim\theta_\lambda$ \citep{Boldyrev2006,Mallet2016_measures_anisotropy_intermittency,Mallet2017_anisotropy}, so the measured $\theta_\lambda\sim\lambda^{1/8}$ gives $\xi\sim\lambda^{7/8}$ rather than $\lambda^{3/4}$. This weaker sheet anisotropy changes the expected onset of a tearing-mediated cascade \citep{Loureiro2017_reconnection_in_turbulence,Mallet2017_plasmoid_disruptions,Boldyrev2017_MHD_mediated_by_reconnection,Comisso2018_MHD_turbulence_plasmoid_regime}. For alignment and fluctuation-amplitude exponents $a$ and $\zeta$, the standard tearing balance gives
    \begin{align}
        k_*L\sim\Rm^{1/(1+2a+\zeta)} .
    \end{align}
    Thus $a=\zeta=1/4$ gives $k_*L\sim\Rm^{4/7}$, whereas the measured $a=1/8$, $\zeta\simeq1/4$ gives $k_*L\sim\Rm^{2/3}$. This shifts the tearing transition to smaller scales and substantially increases the $\Rm$ required to separate it from the resistive scale (Supplemental Material, Sec.~II).

    Sign-mixed helicities also enter large-scale dynamo closures through the turbulent EMF, $\bm{\mathcal E}=\langle\bm{\delta}\bfu\times\bm{\delta}\bfB\rangle$, residual kinetic--current helicity in $\alpha$, and cross helicity in $\Upsilon$ \citep{Pouquet1976_strong_mhd_helical_turbulence,Yokoi2023_cross_helicity_review,Brandenburg2023_turbulent_processes_dynamo,Tripathi2026_upsilon_dynamo}. Here $\alpha$ combines residual kinetic and current helicities, whereas $\Upsilon$ is proportional to cross helicity; their cancellations can therefore differ across averaging scales and local environments. Alfv\'enic alignment suppresses the fluctuating cross product but enhances the scalar cross helicity supplying $\Upsilon$, which can dominate shear-layer dynamos \citep{Tripathi2026_upsilon_dynamo}. It also sets a maximum averaging scale $\lambda_{\rm eq}$ over which the signed helicities cancel, limiting the surviving scale-dependent dynamo contributions.
    
    MMS magnetosheath observations already show analogous relaxation-angle PDFs and cascade transfers \citep{Bandyopadhyay2018_incompressive_energy_transfer_MMS,Bandyopadhyay2020_hall_mhd_cascade_MMS,Pecora2023_relaxation_in_magnetosheath}; measuring all three scale-dependent alignments would extend recent solar-wind $\bfu$--$\bfB$ measurements \citep{Sioulas2024_scale_dependent_alignment_solar_wind}.
\vspace{-5mm}
\section*{Acknowledgments}
\vspace{-2mm}
    J.~R.~B. acknowledges Christoph Federrath, whose help and guidance made it possible to perform the calculations at these scales. We acknowledge productive discussions with Bindesh Tripathi, Elias Most, Ralf S. Klessen, Salvatore Cielo, Uddipan Banik, Riddhi Bandyopadhyay, Eliot Quataert, Sasha Philippov, Philipp K.~S. Kempski, Drummond Fielding, Alexander Chernoglazov, Hui Li, Philip Mocz, Bart Ripperda, and Chris Thompson. J.~R.~B. acknowledges high-performance computing resources provided by the Leibniz Rechenzentrum and the Gauss Centre for Supercomputing grants pr73fi, pn76gi, and pn76ga, and GCS large-scale project 10391. J.~R.~B. and A.~B. acknowledge support from NSF Award 2206756. J.~R.~B. further acknowledges funding from the Natural Sciences and Engineering Research Council of Canada (NSERC) (funding reference number 568580). This work was in part supported by NASA through NASA Hubble Fellowship grant HST-HF2-51597.001-A, awarded by the Space Telescope Science Institute and administered by AURA for NASA under contract NAS5-26555.

\IfFileExists{supplemental_material.bib}
    {\bibliography{beattie,supplemental_material}}
\clearpage
\section*{End Matter}\label{sec:endmatter_start}
\vspace{-2mm}
\subsection{Numerical details and Reynolds estimate}
\vspace{-2mm}
\label{app:reynolds_number}
\label{app:init_conditions}
    To avoid spending the highest-resolution run on transient states, we initialize the largest grids with steady-state realizations from lower-resolution simulations. For both simulations, we drive $\sim 5,\!000^3$ boxes to stationarity, interpolate the steady-state turbulence to the $10,\!000^3$ grids, then drive for two more outer-scale turnover times $2t_0$, and average the main statistics, structure functions and PDFs, over 15 stationary realizations in the final $1.5t_0$. The forcing sets the target rms velocity $\langle u^2\rangle_{\V}^{1/2}$; in the net-flux run we then choose the imposed uniform field $\bfB_0=B_0\hat{\bm z}$ so that $\mathcal{M}_{A0}=\langle u^2\rangle_{\V}^{1/2}/(B_0/\sqrt{\rho_0})=2$, matching the rms Alfv\'enic Mach number of the fluctuation-dynamo simulation.
    The simulations are implicit large-eddy simulations, with effective viscosity and resistivity supplied by the discretization. Calibrations against explicit-viscosity DNS give empirical conversions from grid scale to $\Re$ and $\Rm$ \citep{Kriel2022_kinematic_dynamo_scales,Grete2023_as_a_matter_of_dynamical_range,Shivakumar2023_numerical_dissipation}. For supersonic MHD turbulence, \citet{Shivakumar2023_numerical_dissipation} find $\Re=(N_{\rm grid}/N_{\Re})^{p_{\Re}}$ with $p_{\Re}\in[1.5,2.0]$, $N_{\Re}\in[0.8,4.4]$, and $\Rm=(N_{\rm grid}/N_{\Rm})^{p_{\Rm}}$ with $p_{\Rm}\in[1.2,1.6]$, $N_{\Rm}\in[0.1,0.7]$. For $N_{\rm grid}\simeq10^4$, this gives $\Re\in[1.4\times10^6,5.3\times10^6]$ and $\Rm\in[1.0\times10^6,4.5\times10^6]$.

\vspace{-5mm}
\subsection{Definitions of scale-dependent alignment structure functions}\label{app:scale_dependent_dfn}
\vspace{-2mm}
    To compute the scale-dependent alignment structure functions shown in Fig.~\ref{fig:alignment_variables}(d), we sample point pairs $(\bfx,\bfx+\bfell)$, where $\bfx$ is position and $\bfell$ is the lag vector. The scalar separation $\lambda$ denotes the strong-cascade scale in each geometry. In the no-net-flux run,
    \begin{align}
        \lambda=|\bfell|.
    \end{align}
    The direction of $\bfell$ is not fixed and is not conditioned on the local magnetic field. In the guide-field run, let $\hat{\bfB}_0=\bfB_0/B_0$ be the imposed-field direction and define
    \begin{align}
        \bfell_\perp
        =
        \bfell-(\bfell\cdot\hat{\bfB}_0)\hat{\bfB}_0,
        \qquad
        \lambda=|\bfell_\perp|.
    \end{align}
    In practice, following \citet{Mason2006_dynamical_alignment}, the guide-field lags are sampled in the perpendicular plane, $\bfell\cdot\hat{\bfB}_0=0$, so $\lambda=|\bfell_\perp|=|\bfell|$. For any vector fields $\bm X$ and $\bm Y$, we define raw increments in the $\lambda$ bin as
    \begin{align}
        \bm{\delta}\bm X(\bfx,\bfell) &= \bm X(\bfx+\bfell)-\bm X(\bfx), \\
        \bm{\delta}\bm Y(\bfx,\bfell) &= \bm Y(\bfx+\bfell)-\bm Y(\bfx),
    \end{align}
    with the appropriate $\lambda$ lying in that bin. Following the standard local mean-field definition \citep{Chernoglazov2021_alignment_SR_MHD,Dong2022_reconnection_mediated_cascade}, we define
    \begin{align}
        \widehat{\bfB}_{\bfell}
        =
        \frac{\bfB(\bfx)+\bfB(\bfx+\bfell)}
        {|\bfB(\bfx)+\bfB(\bfx+\bfell)|},
        \qquad
        \P_{\perp,\bfell}
        =
        \mathsf{I}-\widehat{\bfB}_{\bfell}\widehat{\bfB}_{\bfell},
    \end{align}
    and project the increments into the plane perpendicular to this two-point mean magnetic field,
    \begin{align}
        \bm{\delta}\bm X_{\perp,\lambda}
        &=\P_{\perp,\bfell}\bm{\delta}\bm X,\\
        \bm{\delta}\bm Y_{\perp,\lambda}
        &=\P_{\perp,\bfell}\bm{\delta}\bm Y .
    \end{align}
    We then construct the ratio of first-order structure functions
    \begin{align}
    \theta_{\bm X,\bm Y}(\lambda)
    \sim |\sin\theta_{\bm X,\bm Y}(\lambda)|
    = \frac{\Exp{|\bm{\delta}\bm X_{\perp,\lambda}\times\bm{\delta}\bm Y_{\perp,\lambda}|}{\lambda}}
    {\Exp{|\bm{\delta}\bm X_{\perp,\lambda}| |\bm{\delta}\bm Y_{\perp,\lambda}|}{\lambda}}.
    \end{align}
    Here $\Exp{\cdots}{\lambda}$ denotes an average over positions, sampled lag directions, and stationary-state realizations in the separation bin. The no-net-flux construction is applied to the pairs $(\bfu,\bfB)$, $(\bfB,\bfj)$, and $(\bfu,\bfo)$; the guide-field comparison applies the corresponding perpendicular-plane construction to the same three pairs. We restrict attention to this local-field projection because scale-dependent anisotropy and alignment of the strong MHD cascade are recovered in the local mean-field coordinate \citep{Cho2000,Maron2001,Mason2006_dynamical_alignment,Mallet2016_measures_anisotropy_intermittency,Dong2018_role_of_plasmoid_instability,Dong2022_reconnection_mediated_cascade}. Since the no-net-flux run has $\Exp{\bfB}{\V}=0$, it has no global parallel direction to exclude; the isotropic $\lambda=|\bfell|$ statistic is therefore the zero-guide-field analogue of the perpendicular guide-field statistic. Both choices identify the separation across which the strong cascade transfers fluctuation energy. We use $2\times10^{12}$ sampled point pairs to ensure convergence of the structure functions on all scales \citep{Federrath2021,Beattie2025_nature_astro}.

\vspace{-5mm}
\subsection{Cross and kinetic helicity transport in the barotropic and force-free limits}\label{app:barotropic_helicity}
\vspace{-2mm}
    The ideal, barotropic, unforced (inertial) velocity, induction, and vorticity equations are
    \begin{align}
        \partial_t\bfu
        &=\bfu\times\bfo-\bnab\left(\frac{u^2}{2}+h\right)+\bm f_L,\label{eq:app_barotropic_momentum}\\
        \partial_t\bfB
        &=\bnab\times(\bfu\times\bfB),\label{eq:app_barotropic_induction}\\
        \partial_t\bfo
        &=\bnab\times(\bfu\times\bfo)+\bnab\times\bm f_L.
        \label{eq:app_barotropic_vorticity}
    \end{align}
    Here we define the Lorentz acceleration
    \begin{align}
        \bm f_L=\frac{\bfj\times\bfB}{\rho},
    \end{align}
    which vanishes in the Taylor-relaxed limit, $\bfj=\alpha(\bfx)\bfB$, and is gradient-like in the pressure-balanced limit, $\bm f_L\simeq\bnab\chi$ for scalar $\chi$. We use ideal MHD with a barotropic plasma equation of state (including the isothermal equation solved here and general isentropic adiabatic plasmas), so the pressure force is a gradient,
    \begin{align}
        \frac{1}{\rho}\bnab p = \bnab h,
        \qquad
        h(\rho)=\int^\rho \frac{1}{\rho'}\frac{dp}{d\rho'}\,d\rho',
    \end{align}
    which means the pressure term may enter through the conservative fluxes. 
    
    \subsubsection{Helicity evolution equations}    
    Taking $\partial_t(\bfu\cdot\bfB)$ gives
    \begin{align}
        \partial_t H_C+\bnab\cdot\bm F_C=0,
        \label{eq:app_crosshel_balance}
    \end{align}
    with $H_C=\bfu\cdot\bfB$ and
    \begin{align}
        \bm F_C
        =\left(\frac{u^2}{2}+h\right)\bfB+\bfu\times(\bfu\times\bfB), 
    \end{align}
    The Lorentz acceleration drops out identically because $\bfB\cdot\bm f_L=\bfB\cdot(\bfj\times\bfB)/\rho=0$, and the barotropic pressure force contributes as an enthalpy flux. Thus $H_C$ is a compressible, ideal invariant.   
    Taking $\partial_t(\bfu\cdot\bfo)$ gives
    \begin{align}
        \partial_t K+\bnab\cdot\bm F_K =2\bfo\cdot\bm f_L,
        \label{eq:app_kinhel_balance}
    \end{align}
    where $K=\bfu\cdot\bfo$, with flux
    \begin{align}
        \bm F_K =K\bfu+ \left(h-\frac{u^2}{2}\right)\bfo +\bfu\times\bm f_L .
    \end{align}
    In the absence of baroclinicity, $|\bnab\rho \times \bnab p| = 0$, density enters only through the Lorentz acceleration. In the force-free case the kinetic-helicity source vanishes pointwise, while in the gradient-like case $2\bfo\cdot\bm f_L \simeq 2\bfo\cdot\bnab\chi =2\bnab\cdot(\chi\bfo)$, so the source becomes a conservative flux and $\partial_t K+\bnab\cdot\bm F_K \approx 0$. Approximately conserved kinetic-helicity transport therefore requires a small non-gradient part of $(\bfj\times\bfB)/\rho$, and the measured $\bfB$--$\bfj$ alignment is the stronger sufficient condition. Thus, in an ideal, unforced, force-free, barotropic but compressible plasma, both cross helicity, $\bfu\cdot\bfB$, and kinetic helicity, $\bfu\cdot\bfo$, are ideal invariants, justifying the structure functions studied in Fig.~\ref{fig:alignment_variables}. 

\vspace{-5mm}
\subsection{Departure transfer diagnostics}\label{app:alfvenic_transfer}
\vspace{-2mm}
    We use Fourier shell filters $\mathcal{P}_K$, with $\bm f_K=\mathcal{P}_K\bm f$ and $\bm f_{\leq m}=\sum_{P\leq m}\bm f_P$. No-net-flux transfers are binned in $k$; guide-field transfers are binned in $k_\perp$ and summed over $k_\parallel$. For any conservative shell transfer $T_X(K|P)$, the downscale flux through shell boundary $m$ is
    \begin{align}
        \Pi_X(m)
        &=
        \sum_{K>m}\sum_{P\leq m}T_X(K|P).
        \label{eq:app_generic_flux}
    \end{align}
    We write $dV$ for integrals over the periodic volume. Compressible advective transfers use
    \begin{align}
        \mathcal{A}_{\bm c}\bm q_P
        &=
        (\bm c\cdot\bnab)\bm q_P
        +\frac{1}{2}\bm q_P(\bnab\cdot\bm c),
        \label{eq:app_skew_adjoint_operator}
    \end{align}
    with the same definition for scalar $q_P$. This operator is skew-adjoint for periodic boundaries, and the half-divergence term is required for velocity-mediated transfers in compressible flow \citep{Grete2017_shell_models_for_CMHD}. The magnetic field $\bfB$ below is the field used in the alignment statistic and cross helicity, up to a constant code-unit normalization; no local factor of $\rho^{-1/2}$ is applied.

    The total-energy transfer plotted in Fig.~\ref{fig:alfvenic_flux} is
    \begin{align}
        T_E(K|P)
        =&-\int \bfu_K\cdot\mathcal{A}_{\bfu}\bfu_P\,dV
          -\int \bfB_K\cdot\mathcal{A}_{\bfu}\bfB_P\,dV
        \nonumber\\
        &+\int \bfu_K\cdot\mathcal{A}_{\bfB}\bfB_P\,dV
          +\int \bfB_K\cdot\mathcal{A}_{\bfB}\bfu_P\,dV .
        \label{eq:app_energy_transfer}
    \end{align}

    For Alfv\'enic departure, we define a fixed-sign departure and mediator fields
    \begin{align}
        \bm d^\sigma=\bfu-\sigma\bfB,\qquad
        \bm a^\sigma=\bfu+\sigma\bfB,\qquad \sigma=\pm1 .
        \label{eq:app_alfvenic_fields}
    \end{align}
    The fixed-sign shell transfer is
    \begin{align}
        T_{\mathcal{D}_A}^\sigma(K|P)
        &=
        -\int\bm d_K^\sigma\cdot
        \mathcal{A}_{\bm a^\sigma}\bm d_P^\sigma\,dV .
        \label{eq:app_alfvenic_fixed_sign_transfer}
    \end{align}
    Figure~\ref{fig:alfvenic_flux} shows the sign-even matrix $T_{\mathcal{D}_A}=(T_{\mathcal{D}_A}^++T_{\mathcal{D}_A}^-)/2$. For the flux, the nearest local Alfv\'enic branch is chosen at each cutoff,
    \begin{align}
        \sigma_m(\bfx)
        &=
        \operatorname{sgn}\!\left[
        \bfu_{\leq m}(\bfx)\cdot\bfB_{\leq m}(\bfx)
        \right],
        \nonumber\\
        \Pi_{\mathcal{D}_A}(m)
        &=
        -\sum_{\sigma=\pm1}
        \int\mathbb{I}_{\sigma_m=\sigma}
        \bm d_{>m}^{\sigma}\cdot
        \mathcal{A}_{\bm a^\sigma}\bm d_{\leq m}^{\sigma}\,dV .
        \label{eq:app_alfvenic_nearest_flux}
    \end{align}
    The local sign mask makes $\Pi_{\mathcal{D}_A}$ a scale-local flux diagnostic rather than a conventional antisymmetric shell-transfer matrix.

    For the Beltrami departure, we define the velocity component perpendicular to vorticity and the transported vector
    \begin{align}
        \bfu_\perp
        &=
        \bfu-\frac{\bfu\cdot\bfo}{\omega^2}\bfo,
        \nonumber\\
        \bm d_B
        &=
        \bfu_\perp\sin\theta_{\bfu,\bfo},
        \nonumber\\
        \sin^2\theta_{\bfu,\bfo}
        &=
        \frac{|\bfu\times\bfo|^2}{u^2\omega^2},
        \label{eq:app_beltrami_vector_departure}
    \end{align}
    with $\bm d_B=0$ where $u^2\omega^2$ is numerically zero, and $u=|\bfu|$, $\omega=|\bfo|$. Here $\bfu_\perp$ is the velocity component that vanishes in either Beltrami branch, while the extra factor of $\sin\theta_{\bfu,\bfo}$ makes $|\bm d_B|^2=u^2\sin^4\theta_{\bfu,\bfo}$, the amplitude-weighted, sign-even second-moment departure used above. Its conservative transfer and flux are
    \begin{align}
        T_{\mathcal{D}_B}(K|P)
        &=
        -\int\bm d_{B,K}\cdot
        \mathcal{A}_{\bfu}\bm d_{B,P}\,dV ,
        \nonumber\\
        \Pi_{\mathcal{D}_B}(m)
        &=
        \sum_{K>m}\sum_{P\leq m}
        T_{\mathcal{D}_B}(K|P).
        \label{eq:app_beltrami_flux}
    \end{align}
    By construction, $T_{\mathcal{D}_B}(K|P)=-T_{\mathcal{D}_B}(P|K)$.

\clearpage
\onecolumngrid
\setcounter{page}{1}
\renewcommand{\thepage}{S\arabic{page}}
\begin{center}
    {\large\bfseries Supplemental Material for\\[0.4em]
    Local relaxation and scale-dependent alignment\\
    in compressible, magnetized turbulence\par}
    \medskip
    James R. Beattie and Amitava Bhattacharjee
    \medskip

    {\small Citation numbers refer to the reference list of the accompanying Letter.}
\end{center}
\label{sec:supplemental_start}
\twocolumngrid
\setcounter{section}{0}
\setcounter{figure}{0}
\setcounter{equation}{0}
\setcounter{table}{0}
\renewcommand{\thesection}{\Roman{section}}
\renewcommand{\thefigure}{S\arabic{figure}}
\renewcommand{\theequation}{S\arabic{equation}}
\renewcommand{\thetable}{S\arabic{table}}
\section*{I. Visualizations of local relaxed-state alignment}
\begin{figure*}
    \centering
    \includegraphics[width=\textwidth]{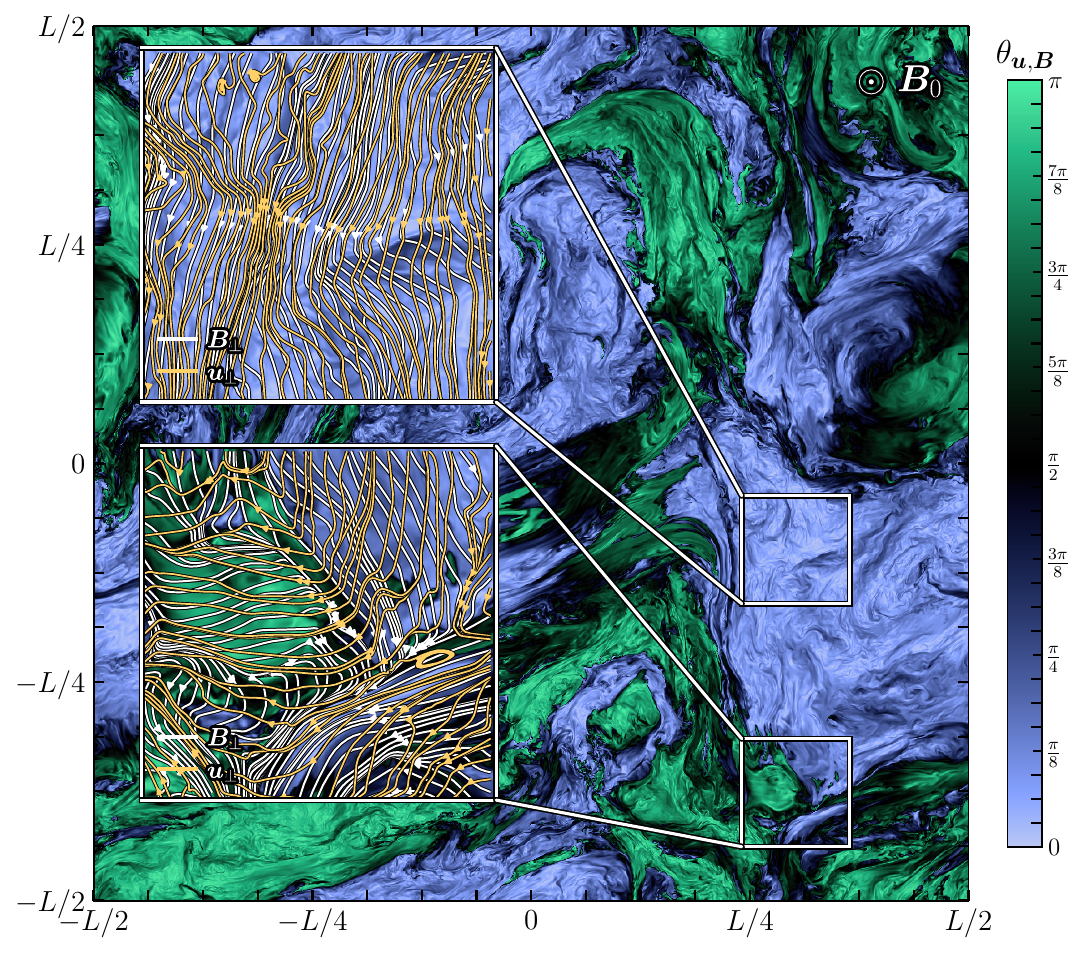}
    \caption{\textbf{Guide-field Alfv\'enic alignment.} The color shows $\theta_{\bfu,\bfB}$ on an $xy$ slice perpendicular to the imposed guide field, $\bfB_0$, in the net-flux simulation. The white boxes mark a strongly aligned patch and a nearby interface between aligned patches. The insets show the same regions with in-plane magnetic-field streamlines, $\bfB_\perp$, in white and in-plane velocity streamlines, $\bfu_\perp$, in gold.}
    \label{fig:supp_alignment_zoom}
\end{figure*}

\vspace{-2mm}
    Fig.~\ref{fig:supp_alignment_zoom} shows the guide-field simulation in real space, using the same $\theta_{\bfu,\bfB}$ angle as in the main text. Most of the slice is close to one of the two Alfv\'enic branches, $\bfu\parallel\bfB$ or $\bfu\parallel-\bfB$, while the dark, nearly orthogonal regions are comparatively thin interfaces where the Alfv\'enic nonlinearity is largest. In this sense, the visualization suggests that Alfv\'enic alignment is the volume-filling organization, with misalignment concentrated at boundaries between sign-mixed relaxed patches. The two zooms show an aligned region and a nearby interface; in both cases, the in-plane velocity and magnetic streamlines remain locally organized, consistent with the local-relaxation picture.

\begin{figure*}
    \centering
    \includegraphics[width=\textwidth]{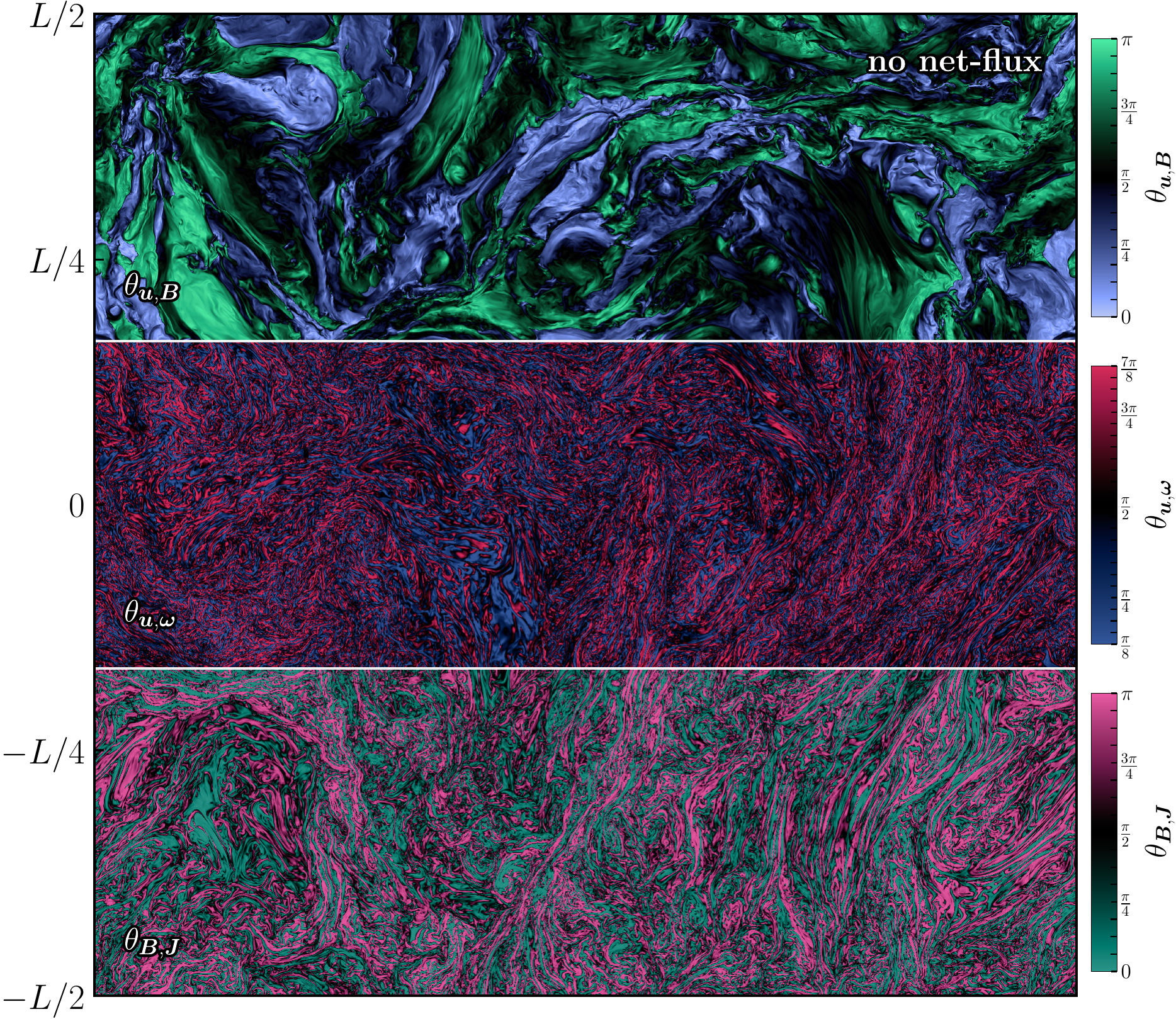}
    \caption{\textbf{Three relaxed-state alignment angles.} Composite no-net-flux slice showing the Alfv\'enic angle $\theta_{\bfu,\bfB}$ in the top partition, the Beltrami angle $\theta_{\bfu,\bfo}$ in the middle partition, and the Taylor angle $\theta_{\bfB,\bfj}$ in the bottom partition. The color encodes the local angle from parallel through perpendicular to anti-parallel alignment; the $\theta_{\bfu,\bfo}$ color scale is clipped at $\pi/8$ and $7\pi/8$ for contrast.}
    \label{fig:supp_threeway_alignment}
\end{figure*}

    Fig.~\ref{fig:supp_threeway_alignment} shows the corresponding full-slice organization of the three alignment angles used in the Letter. The Alfv\'enic, Beltrami, and Taylor angles all show extended aligned or anti-aligned regions separated by sharper, nearly orthogonal interfaces, reinforcing the interpretation that the measured structure functions quantify departures from locally relaxed states.

\vspace{-5mm}
\section*{II. Critical length scale for tearing-mediated cascade}
\vspace{-2mm}
    Starting from the disruption criterion used in reconnection-mediated cascade models \citep{Loureiro2017_reconnection_in_turbulence,Mallet2017_plasmoid_disruptions,Boldyrev2017_MHD_mediated_by_reconnection,Comisso2018_MHD_turbulence_plasmoid_regime}, the critical eddy satisfies, up to the logarithmic Lambert-$W$ factor and order-unity constants,
    \begin{align}
        \frac{\lambda_*}{\xi_*}
        \sim
        S_{\xi_*}^{-1/3},
        \qquad
        S_{\xi_*}
        =
        \frac{\xi_*\delta b_{\lambda_*}}{\eta}.
        \label{eq:supp_tearing_balance_general}
    \end{align}
    Here $\lambda_*$ and $\xi_*$ are the reconnecting-sheet thickness and length, $S_{\xi_*}$ is the Lundquist number based on $\xi_*$, $\delta b_{\lambda_*}$ is the magnetic-fluctuation amplitude at $\lambda_*$, and $\eta$ is the magnetic diffusivity. Let $a$ be the alignment exponent and $\zeta$ the magnetic-amplitude exponent, then,
    \begin{align}
        \frac{\lambda}{\xi}\sim\theta_\lambda\sim
        \left(\frac{\lambda}{L}\right)^a,
        \qquad
        \delta B_\lambda\sim
        \left(\frac{\lambda}{L}\right)^\zeta .
    \end{align}
    The aligned eddy length $\xi_\lambda$ and corresponding Lundquist number $S_\xi$ are then
    \begin{align}
        \xi_\lambda\sim L\left(\frac{\lambda}{L}\right)^{1-a},
        \qquad
        S_{\xi}\sim
        \Rm\left(\frac{\lambda}{L}\right)^{1-a+\zeta}.
    \end{align}
    Substituting into Eq.~\eqref{eq:supp_tearing_balance_general} gives
    \begin{align}
        \left(\frac{\lambda_*}{L}\right)^a
        \sim
        \Rm^{-1/3}
        \left(\frac{\lambda_*}{L}\right)^{-(1-a+\zeta)/3},
    \end{align}
    or
    \begin{align}
        \frac{\lambda_*}{L}
        \sim
        \Rm^{-1/(1+2a+\zeta)},
        \qquad
        k_*L\sim
        \Rm^{1/(1+2a+\zeta)}.
        \label{eq:supp_critical_scale_general}
    \end{align}
    where $k_*\sim1/\lambda_*$ is the critical wavenumber. For the usual dynamic-alignment scaling $a=\zeta=1/4$, this recovers
    \begin{align}
        k_*L\sim\Rm^{4/7}.
    \end{align}
    For the present alignment law, $a=1/8$, together with the measured subsonic $k^{-3/2}$ velocity scaling $\zeta\simeq1/4$, Eq.~\eqref{eq:supp_critical_scale_general} gives
    \begin{align}
        \frac{\lambda_*}{L}\sim\Rm^{-2/3},
        \qquad
        k_*L\sim\Rm^{2/3}.
    \end{align}
    Thus the weaker Alfv\'enic alignment shifts the tearing-mediated transition to smaller scales than the standard dynamic-alignment estimate. The same scaling also changes the Reynolds number needed for a resolved tearing-mediated interval. If the onset criterion is a fixed separation between the disruption scale and the resistive scale,
    \begin{align}
        \frac{\lambda_\eta}{L}
        \sim
        \Rm^{-1/(1+\zeta)},
    \end{align}
    then
    \begin{align}
        \frac{\lambda_*}{\lambda_\eta}
        \sim
        \Rm^{1/(1+\zeta)-1/(1+2a+\zeta)} .
        \label{eq:supp_tearing_resolved_interval}
    \end{align}
    For $\zeta=1/4$, the exponent in Eq.~\eqref{eq:supp_tearing_resolved_interval} changes from $8/35$ when $a=1/4$ to $2/15$ when $a=1/8$. Therefore a critical value inferred in the standard dynamic-alignment case maps to
    \begin{align}
        \Rm_c(a=1/8)
        \sim
        \left[\Rm_c(a=1/4)\right]^{12/7}.
    \end{align}
    Taking $\Rm_c(a=1/4)\sim10^5$, consistent with the threshold discussed by \citet{Dong2022_reconnection_mediated_cascade}, gives $\Rm_c(a=1/8)\sim4\times10^8$, up to order-unity and logarithmic corrections.

\vspace{-5mm}
\section*{III. PVNLT relaxed states}\label{app:pvnlt}
\vspace{-2mm}
    For an inviscid invariant built from fields $\bm p$ and $\bm q$,
    \begin{align}
        M=\int \bm p\cdot\bm q\,d^3x
    \end{align}
    PVNLT \citep{Banerjee2023_relaxed_states} considers the two-point correlator at separation $\bfell$,
    \begin{align}
        R_M(\bfell)
        =\left\langle
        \frac{\bm p(\bfx)\cdot\bm q(\bfx+\bfell)
        +\bm p(\bfx+\bfell)\cdot\bm q(\bfx)}{2}
        \right\rangle,
    \end{align}
    where the average is over position $\bfx$. The schematic evolution of the correlator is
    \begin{align}
        \partial_t R_M
        =\left\langle F_{\rm tr}^M\right\rangle
        +\left\langle f_c^M\right\rangle
        +\left\langle d_c^M\right\rangle .
    \end{align}
    Here $F_{\rm tr}^M$ is the nonlinear transfer of the correlator, while $f_c^M$ and $d_c^M$ denote forcing and dissipation contributions. In a stationary inertial range, $\partial_t R_M=0$ and $\left\langle f_c^M\right\rangle=\left\langle d_c^M\right\rangle=0$, so the nonlinear transfer term must vanish,
    \begin{align} \label{eq:relaxed_transfers}
        \left\langle F_{\rm tr}^M\right\rangle=0 .
    \end{align}
    This is the condition for relaxation in PVNLT. Applying Eq.~\eqref{eq:relaxed_transfers} to incompressible MHD gives
    \begin{align}
        \bfu\times\bfo+\bfj\times\bfB &= \bnab(P_T+\phi_0),\label{eq:pvnlt_mhd1}\\
        \bfu\times\bfB &= \bnab\psi_0,\label{eq:pvnlt_mhd2}
    \end{align}
    for the energy, magnetic-helicity, and cross-helicity correlators, where $P_T=p+u^2/2$ is the total pressure built from plasma pressure $p$ and kinetic pressure $u^2/2$, and $\phi_0,\psi_0$ are scalar fields \citep{Banerjee2023_relaxed_states}. Following \citet{Banerjee2023_relaxed_states}, the aligned limits follow by taking the scalar-gradient part of the induction condition to be negligible. If $\bnab\psi_0\simeq0$, then
    \begin{align}
        \bfu\times\bfB\simeq0
        \quad\Rightarrow\quad
        \bfu=\kappa\bfB ,
        \qquad
        \bfo=\kappa\bfj,
    \end{align}
    where the last relation follows for constant $\kappa$. Substitution into Eq.~\eqref{eq:pvnlt_mhd1} gives
    \begin{align}
        (1-\kappa^2)\bfj\times\bfB
        =
        \bnab(P_T+\phi_0).
        \label{eq:pvnlt_taylor_limit}
    \end{align}
    Thus the relaxed state is generally pressure- or gradient-balanced. In the low-$\beta$, magnetically dominated limit discussed by \citet{Banerjee2023_relaxed_states}, Eq.~\eqref{eq:pvnlt_taylor_limit} reduces to $\bfj\times\bfB\simeq0$, the Taylor or force-free limit. If instead $\kappa=\pm1$, the same branch gives $\bfu=\pm\bfB$ and Eq.~\eqref{eq:pvnlt_mhd1} requires only a constant total pressure, the Alfv\'enic limit. Finally, Eq.~\eqref{eq:pvnlt_mhd1} shows the Beltrami limit: if the Lorentz term is force-free or pressure-balanced, then the non-gradient part of $\bfu\times\bfo$ is depleted, and the negligible-gradient limit gives $\bfu\times\bfo\simeq0$. Thus PVNLT \citep{Banerjee2023_relaxed_states} naturally yields pressure- or gradient-balanced relaxed states, with Taylor, Alfv\'enic, and Beltrami alignments as measurable limiting cases.

\vspace{-5mm}
\section*{IV. Estimate for the alignment exponents}\label{app:alignment_exponents}
\vspace{-2mm}
    The departure model keeps the PVNLT \citep{Banerjee2023_relaxed_states} idea that relaxation depletes nonlinear invariant transfer, but applies it at finite distance from a relaxed state. Let $\delta u_\lambda\sim\delta B_\lambda\sim\lambda^\zeta$ with $\zeta\simeq1/4$ in the measured subsonic range. For an Alfv\'enic state, the normalized positive distance from the nearest $\bfu=\pm\bfB$ state is
    \begin{align}
        \D_A(\lambda)=
        \min_{\sigma=\pm1}
        \frac{|\bm{\delta}\bfu_\lambda-\sigma\bm{\delta}\bfB_\lambda|^2}
        {|\bm{\delta}\bfu_\lambda|^2+|\bm{\delta}\bfB_\lambda|^2}
        \simeq \frac{1}{2}\theta_{\bfu,\bfB}^2(\lambda),
    \end{align}
    for nearly equipartitioned, small-angle increments. If this first-moment departure is carried downscale at a scale-independent rate,
    \begin{align}
        \varepsilon_{\D_A}(\lambda)\sim \frac{\delta u_\lambda^3\D_A(\lambda)}{\lambda},
    \end{align}
    then
    \begin{align}
        \theta_{\bfu,\bfB}(\lambda)\propto\lambda^{(1-3\zeta)/2}\simeq\lambda^{1/8},
    \end{align}
    which gives the measured Alfv\'enic exponent. For the Beltrami departure, the local invariant density is kinetic helicity, $\bfu\cdot\bfo$. It is signed and globally canceled in the present nonhelical system, so the leading sign-even constraint is a second-moment departure. With
    \begin{align}
        \D_B(\lambda)=
        1-\frac{|\bm{\delta}\bfu_\lambda\cdot\bm{\delta}\bfo_\lambda|}
        {|\bm{\delta}\bfu_\lambda||\bm{\delta}\bfo_\lambda|}
        \simeq\frac{1}{2}\theta_{\bfu,\bfo}^2(\lambda),
    \end{align}
    and a scale-independent rate for carrying this second moment downscale,
    \begin{align}
        \varepsilon_{\D_B}(\lambda)\sim \frac{\delta u_\lambda^3\D_B^2(\lambda)}{\lambda},
    \end{align}
    one obtains
    \begin{align}
        \theta_{\bfu,\bfo}(\lambda)\propto\lambda^{(1-3\zeta)/4}\simeq\lambda^{1/16}.
    \end{align}
    The different powers follow from the invariant being transported: the Alfv\'enic alignment type has a positive first-moment distance built from energy and cross helicity, while the Beltrami alignment type is signed and therefore enters through a sign-even second moment. The End Matter section ``Departure transfer diagnostics'' gives the corresponding shell-transfer diagnostics.

\end{document}